\documentstyle[multicol,aps,prl,epsf]{revtex}

\draft

\begin{document}
\draft
\title{Multistability and Critical Fluctuation in a Split Bose-Einstein Condenstate}
\author{Ying $Wu^{1}$ and Chang-pu $Sun^{2}$}
\address{$ ^{1}$ Physics Department, Huazhong University of Science and
Technology, Wuhan 430074, and  Wuhan Institute of Physics and
Mathematics, The Chinese Academy of Sciences, People's Republic of
China}
\address{$^{2}$ Institute of Theoretical Physics, The Chinese Academy of Sciences,
Beijing 100080, People's Republic of China}

\date{\today}
\maketitle
\begin{abstract}
By using a two-mode description, we show that there exist the
multistability, phase transition and associated critical
fluctuations in the macroscopic tunneling process between the
halves of a double-well trap containing a Bose-Einstein
condenstate.  The phase transition that two of the triple stable
states and a unstable state merge into one stable state or a
reverse process takes place whenever the ratio of the mean field
energy per particle to the tunneling energy goes across a critical
value of order one. The critical fluctuation phenomenon
corresponds to squeezed states for the phase difference between
the two wells accompanying with large fluctuations of atom
numbers.
\end{abstract}
\begin{multicols}{2}
Over the last few years, there has been considerable interest in
the subjects related to the macroscopic tunneling processes
between the halves of a double-well trap containing a
Bose-Einstein condenstate (BEC). Here we only mention a few ones
closely related to the topic concerned in the present paper, for
instance, experimentally realized squeezed states\cite{orz}, the
scheme for demonstrating nonlinear Josephson-type oscillations of
a driven, two-component BEC \cite{hol}, coherent oscillations
concerning Josephson effects, $\pi$ oscillations, and macroscopic
quantum self-trapping \cite{rag,sme}. Javanainen and Ivanov
\cite{jav1} have claimed that these studies \cite{hol,rag,sme} are
based on essentially classical models. On the other hand, the
fluctuations of atom numbers and phases have been intensively
investigated quantum-mechanically mainly by two seemingly quite
different approaches: two-mode approximation \cite{jav1,jav,jav2}
and the one based on taking atom number and phase difference as
conjugate quantum variables \cite{com1,leg,sols}. The
corresponding investigations in an array of traps containing BEC
have also been carried out recently \cite{orz,jav}. Although
extensively studied, there still exist some important open issues
in the the macroscopic tunneling processes between the halves of a
double-well trap containing a BEC. Some of these open issues are
1) the fluctuations of atom number and phase difference obtained
by the two above mentioned approaches seem to show large
difference\cite{com1,com2}; 2)there exists no any bridge to
connect the strong and weak tunneling regimes yet \cite{com2}; 3)
how to relate the essentially classical models\cite{rag,sme} and
the corresponding quantum-mechanical description in dealing with
the fluctuations. In this paper, we shall solve these three
important issues by providing a united approach. What is more
important, we shall show that the previous studies have missed the
phenomena of the multistability, phase transition and associated
critical fluctuations in the macroscopic tunneling processes
between the halves of a double-well trap containing a BEC. The
critical fluctuation phenomenon corresponds to the squeezed states
for the phase difference between the two wells with extremely
large fluctuation of atom numbers.

We consider a model system of the many-atom ground state of a
Bose-Einstein condensate in a double-well potential
\cite{orz,hol,rag,sme,jav1,jav,jav2,ima,fis,mil1,ste,yang}. In
this system, $N$ bosonic atoms are confined by a infinite harmonic
potential that is divided into left and right wells by a barrier
that can be raised and lowered arbitrarily. Making a simple
two-mode approximation, considering only the lowest energy states,
the creation and annihilation operators ($\hat{a} ^{\dag}_{L,R}$
and $\hat{a}_{L,R}$, respectively) for atoms localized in the
ground state of either the left or the right potential well can be
constructed. Neglecting terms that depend only on the total
conserved particle number $N$, the Hamiltonian for the system can
be written \cite{orz}
\begin{equation}
\hat{H}=\frac{\Delta}{2}(\hat{n}_{L}-\hat{n}_{R})
+\frac{g\beta}{2}(\hat{n}_{L}^{2}+\hat{n}_{R}^{2})+\gamma(\hat{a}
^{\dag}_{L}\hat{a}_{R}+\hat{a} ^{\dag}_{R}\hat{a}_{L}),
\end{equation}
where $\hat{n}_{L,R}\equiv\hat{a}^{\dag}_{L,R}\hat{a}_{L,R}$, the
total conserved particle number operator
$\hat{N}=\hat{n}_{L}+\hat{n}_{R}$, $g=4 \pi a_{sc}\hbar^{2}/m$ is
the mean-field enegy constant ($a_{sc}$ is the s-wave scattering
length). The term in $\gamma$ describes tunneling between wells,
whereas the term in $g\beta$, which depends on the number of atoms
within each wells, describes mean field energy due to interactions
between atoms in the same well. The term in
$\Delta=(E_{L}-E_{R})/\hbar$ describes the energy difference of
the ground states in the left and right wells. The coefficients
$\gamma$ and $\beta$ are determined from integrals over
single-particle wave functions \cite{orz,jav1}.

Introducing phase operators $\hat{\phi}_{L,R}$ by the relation
\cite{mil,scu,sus,peg}
$\exp(i\hat{\phi}_{L,R})=(\hat{n}_{L,R}+1)^{-1/2}\hat{a}_{L,R}$ or
\begin{equation}
\hat{a}_{L,R}=\exp(i\hat{\phi}_{L,R})\sqrt{\hat{n}_{L,R}}
\end{equation}
where we have made use of the fact
$\hat{a}_{L,R}F(\hat{n}_{L,R})=F(\hat{n}_{L,R}+1)\hat{a}_{L,R}$.
The phase and atom number operators satisfy the commutative
relations,
\begin{equation}
\left[\hat{n}_{L,R},\;\hat{\phi}_{L,R}\right]=i
\end{equation}
But phase operators thus introduced suffer from the well-known
non-Hermitian problem \cite{mil,scu,sus,peg} that
$(\exp(i\hat{\phi}_{L,R}))^{\dag}\exp(i\hat{\phi}_{L,R})=1-|0\rangle\langle
0|\not=1$ although
$\exp(i\hat{\phi}_{L,R})(\exp(i\hat{\phi}_{L,R}))^{\dag}=1$.
However, close inspection of this problem, we realize that this
problem is in fact avoidable if we focus on system's states with
$n_{L,R}\not=0$, which will be assumed to be so hereafter. Let
$\hat{n}=\hat{n}_{L}$ and
$\hat{\phi}=\hat{\phi}_{L}-\hat{\phi}_{R}$ denote the left well's
atom number operator and  the phase difference between the left
and right wells respectively. Utilizing
$\hat{n}_{R}=\hat{N}-\hat{n}$, and the total atom number
$\hat{N}=N$ (the conserved operator $\hat{N}$ only takes a unique
eigenvalue $N$ and hence we need not to consider its operator
characteristic), we can, after omitting the unimportant conserved
quantity $N(\Delta+g \beta N)/2$, rewrite the Hamiltonian (1) as
follows,
\begin{equation}
\hat{H}=\Delta\hat{n} -g\beta\hat{n}(N-\hat{n})
+\gamma\left(\sqrt{\hat{n}(N-\hat{n})}\exp(i\hat{\phi})+h.c.\right)
\end{equation}
where the left well's atom number operator $\hat{n}$ and the
phase-difference operator $\hat{\phi}$ satisfy the commutative
relation
\begin{equation}
\left[\hat{n},\;\hat{\phi}\right]=i
\end{equation}
Equations (4) and (5) are the fully quantum-mechanical model give
the united description for investigating the various problems in a
Bose-Einstein condensate in a double-well potential, particularly
those relevant to the phase coherence. In particular, we shall
demonstrate that all the previous models, either essentially
classical or the fully quantum-mechanical ones, dealing with this
systems can be derived from these equations under some
approximations.

Suppose that the system can be described by a two-mode coherent
state $|\Psi\rangle=|\alpha_{L},\alpha_{R}\rangle_{c}$
characterized by two complex parameters
$\alpha_{L}=\sqrt{n}\exp(i\phi_{L})$ and
$\alpha_{R}=\sqrt{N-n}\exp(i\phi_{R})$, we can derive from (4)
Hamiltonian formalism for the equations of motion for the mean
atom number $n=\langle\Psi|\hat{n}|\Psi\rangle$ in the left well
(as well as the mean atom number $N-n$ in the right well) and
phase difference $\phi=\langle\Psi|\hat{\phi}|\Psi\rangle$ between
the two wells as follows (obtained by taking average operations to
the Heisenberg equation of motion $d \hat{A}/dt =i
[\hat{H},\hat{A}]$ for $\hat{A}=\hat{n}$ and $\hat{\phi}$
respectively),
\begin{equation}
\frac{dn}{dt}=\frac{\partial \mathcal{H}}{\partial \phi},\;\;
\frac{d\phi}{dt}=-\frac{\partial \mathcal{H}}{\partial n}
\end{equation}
\begin{equation}
{\mathcal{H}}={\mathcal{H}}_{0}+\Delta n-g \beta n (N-n)
+2\gamma\sqrt{n(N-n)}\cos \phi
\end{equation}
where ${\mathcal{H}}\equiv\langle\Psi|\hat{H}|\Psi\rangle$, and
${\mathcal{H}}_{0}=N(\Delta+g \beta N)/2$ is conserved quantity
and can be omitted without loss of generality. The explicit form
of equation(6) reads
\begin{mathletters}
\begin{equation}
\frac{dn}{dt}=2\gamma \sqrt{n(N-n)}\sin \phi
\end{equation}
\begin{equation}
\frac{d\phi}{dt}=\Delta-g \beta (N-2n)+\gamma
\frac{N-2n}{\sqrt{n(N-n)}}\cos \phi
\end{equation}
\end{mathletters}
These are nonlinear versions of the usual Josephson-junction
equations \cite{hol} and are nearly identical to those describing
the double-well tunneling problem in the same system
\cite{rag,sme}. Raghavan {\it et al.} have investigated Josephson
effects, $\pi$ oscillations and macroscopic quantum self-trapping
for this system \cite{rag}. However, no one, to the best of our
knowledge, seems to have so far noticed the important phenomena of
multistability and phase transition in the macroscopic tunneling
process between the halves of a double-well trap containing a
Bose-Einstein condenstate, which we now turn to investigate.

The multistability and phase transition in the macroscopic
tunneling process are clearly seen from Fig. 1. Let us describe
their main features.  First of all, there are two kinds of
evolution pattern for atom number $n$ ($N-n$) in the left (right)
well and phase difference between the two wells. They are: 1)
stable and unstable steady states (fixed points in the phase
diagrams) denoting there exists no tunneling at all although
tunneling rate is non-zero. 2) periodic tunneling processes where
the tunneling amplitude can be very large,i.e., macroscopic
quantum tunneling, even for small tunneling rate $\gamma$.
Secondly, the phase transition takes place when an "order"
parameter $|\xi|=|g\beta N /(2\gamma)|$ characterizing the
relative magnitude of the mean field energy per particle and the
tunneling energy goes across the critical parameter $\xi_{c}$ of
order one. In other words, one of the three stable fixed point
remains while the unique unstable fixed point and the two of the
three stable fixed points for $|\xi|>\xi_{c}$ merge into one
stable fixed point when $|\xi|$ goes from below to above the
critical value $\xi_{c}$. Therefore the phase transition
corresponding to the sudden structural change in $n-\phi$ phase
diagrams when the "order" parameter $|\xi|$ goes across its
critical value $\xi_{c}$. Thirdly, There exist three stable and
one unstable fixed points when $|\xi|>\xi_{c}$,whereas there exist
two stable fixed points and no unstable one otherwise. The
concrete value of the critical parameter depends on the parameter
$\delta=\Delta /(2\gamma)$ corresponding to the ratio of the
ground energy difference of the two wells to the tunneling rate.
When $|\xi|>\xi_{c}$, the two of the three stable fixed points
(i.e., the two stable fixed points at $\phi=0$ in the lower phase
diagrams of Fig.1) are symmetric about $n=N/2$ if the right and
left wells are identical to each other ($\delta=0$) and they are
asymmetric if the two wells are different from each other
($\delta\not=0$).
\begin{figure}[btp]
\begin{tabbing}
\=\epsfxsize=8.8cm\epsffile{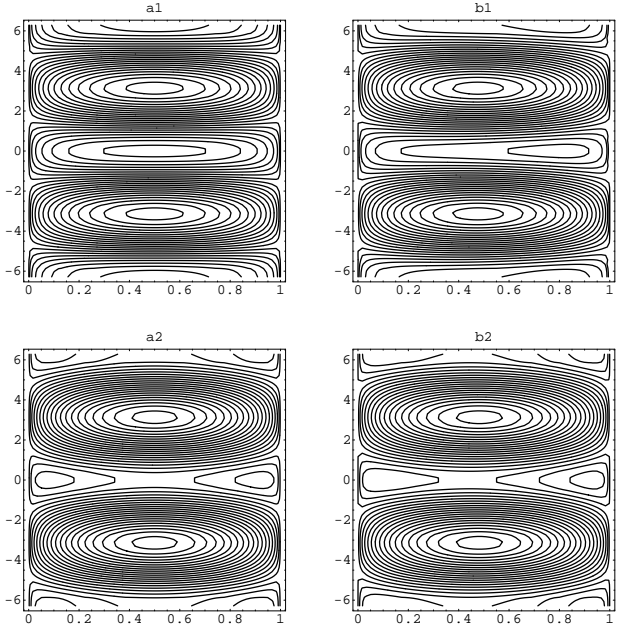}
\end{tabbing}
\caption{The ${\mathcal{H}}/(2\gamma)$ contours for several values
of parameters $\xi=g\beta N /(2\gamma)$ and $\delta =\Delta
/(2\gamma)$. The horizontal and vertical axes denote reduced atom
number $x =n/N$ and the phase difference $\phi$ respectively. The
Parameters for the four diagrams are (a1)$(\xi=0.6,\delta=0)$,
(a2) $(\xi=1.8,\delta=0)$, (b1) $(\xi=1.1,\delta=0.1)$ and (b2)
$(\xi=1.8,\delta=0.1)$ respectively.}
\end{figure}

The fixed points and the critical parameter $\xi_{c}$ are easily
shown to be determined by the equations $\sin \phi_{0} =0
\;\;or\;\phi_{0}=0,\pi$ and
\begin{equation}
(1-2x_{0})\left(\cos\phi_{0}
-2\xi\sqrt{x_{0}(1-x_{0})}\right)=\delta\sqrt{x_{0}(1-x_{0})},
\end{equation}
where $\xi=g\beta N /(2\gamma)$, $\delta =\Delta /(2\gamma)$, and
$x_{0} =n_{0}/N$. We find from this equation that the critical
parameter $\xi_{c}=1$ for $\delta=0$, and $\xi_{c}>1$ for nonzero
$\delta$. For instance, $\xi_{c}\approx 1.37$ for $\delta=0.1$. In
the case $\delta=0$ denoting identical ground energy for two
wells, we can easily obtain the explicit expressions for the fixed
points. They are the fixed point $P$ whose $n_{0}=N/2$ and
$\phi_{0}=0 $ for $\xi>0$ and $\phi_{0}=\pi$ for $\xi<0$, the
fixed points $S_{\pm}$ whose $n_{0}=N(1\pm\sqrt{1-\xi^{-2}})/2$
and $\phi_{0}=0 $ for $\xi>0$ and $\phi_{0}=\pi$ for $\xi<0$, the
fixed point $S$ whose $n_{0}=N/2$ and $\phi_{0}=\pi $ for $\xi>0$
and $\phi_{0}=0$ for $\xi<0$. The fixed point $P$ is unstable for
$|\xi|>1$ and stable for $|\xi| \le 1$, $S$ is stable and exists
for any $\xi$, and $S_{\pm}$ are stable but they exist only when
$|\xi|\ge 1$.

Now let us illustrate that (4) and (5) provide a natural basis for
describing quantum-mechanically the atom number and phase
statistics.  To this goal, we expand the left well's atom number
operator $\hat{n}$ and the phase-difference operator $\hat{\phi}$
in the Hamiltonian (4) around one of the stable steady states
discussed in the paragraph where (9) locates, i.e.,
$\hat{\phi}=n_{0}+\hat{\eta}$ and $\hat{\phi}=\phi_{0}+\hat{\psi}$
with $(n_0,\phi_{0})$ denoting one of the stable steady states
(note $\sin\phi_{0}=0$) and $[\hat{\eta},\;\hat{\psi}]=i$. Then
after neglecting the terms equal to or higher than the order of
${\mathcal{O}}\left(
\hat{\eta}^{3},\hat{\eta}\hat{\psi}^{2},\hat{\eta}^{2}\hat{\psi},
\hat{\psi}^{3}\right)$ and taking $\hat{\psi}=-i
\partial/\partial \eta$ and $\hat{\eta}=\eta$ in order to satisfy
$[\hat{\eta},\;\hat{\psi}]=i$, we can write (4) as follows,
\begin{equation}
\hat{H}\approx
H_{0}+E_{1}\frac{\partial}{\partial\eta}+E_{2}\eta\frac{\partial}{\partial\eta}
-\frac{E_{J}}{2}\frac{\partial^{2}}{\partial\eta^{2}}+\frac{E_{C}}{2}\eta^{2}
\end{equation}
where $H_{0}$ is a constant equal to
$({\mathcal{H}}-{\mathcal{H}}_{0})$ in (7) evaluated at the fixed
point considered, and the coefficients in (10) are given in (11)
for the fixed point $S$ and in (13) for the fixed points $S_{\pm}$
respectively. Following the same argument as the one in
ref.\cite{jav1} given in the paragraph immediately after its
equation (16) , we can neglected $E_{1,2}$ terms in (10) in
evaluated the atom number and phase statistics. After this
approximation, (10) is nothing but a harmonic oscillator model and
fluctuations of the atom number in the left well and the phase
difference are easily shown to be $\Delta n=(\Delta
\phi)^{-1}=(E_{J}/E_{C})^{1/4}$ when the harmonic oscillator is in
its ground state.

In the case where $\delta=0$ denoting identical ground energy for
the two wells, if one considers the stable fixed point $S$ whose
concrete expression is given after (9), the coefficients in (10)
can be calculated to be
\begin{equation}
E_{1}=0, \;E_{2}=\frac{g\beta}{|\xi|},\;E_{J}=\frac{g\beta
N^{2}}{2|\xi|}, \;E_{C}=\frac{2g\beta}{1+|\xi|}.
\end{equation}
Consequently, fluctuations of the atom number in the left well and
the phase difference in this case are given by
\begin{equation}
\Delta n=\frac{1}{\Delta
\phi}=\frac{\sqrt{N}}{\sqrt{2}(1+|\xi|)^{1/4}}
\end{equation}
where $\xi=g\beta N /(2\gamma)$. The equation (16) of
ref.\cite{jav1} dealing with the same system as ours has nearly
identical form as our (10) and (11) but with two slightly
different coefficients $ E_{C}=2g\beta (1+0.5|\xi|^{-1})$ and
$E_{2}=0.5 g\beta/|\xi|$ in our notation. It is instructive to
note that the atom number fluctuation $\Delta n \propto N^{1/4}$
in the weak tunneling regime ($|\xi|\gg 1$) just as given by
Leggett and Sols \cite{com1}, whereas in the strong tunneling
regime ($|\xi|\ll 1$), the atom number fluctuation $\Delta n
\propto \sqrt{N}$ just as obtained by Javanainen and Wilkins
\cite{jav2,com2}. Therefore the problem quarreled by them
\cite{com1,com2} are naturally settled.

The previous studies on the atom number statistics for the
double-well system have failed to notice the multistability and
the phase transition in this system when the parameter $|\xi|$
goes across the critical parameter $\xi_{c}$. Consequently, no one
has so far discussed the atom number and phase statistics around
the stable fixed points $S_{\pm}$ when $|\xi|> \xi_{c}$, which
will be the subject of the present paragraph. The critical
parameter $\xi_{c}$ is unity in the case where $\delta=0$ denoting
identical ground energy for the two wells. In this case, we
consider fluctuations around the stable fixed points $S_{\pm}$
whose concrete expression is given after (9), and obtain the
coefficients in (10) as follows,
\begin{mathletters}
\begin{equation}
E_{1}=\pm 0.5 g\beta N \sqrt{1-\xi^{-2}}, \;E_{2}=-g\beta\xi^{2},
\end{equation}
\begin{equation}
E_{J}=-\frac{g\beta N^{2}}{2\xi^{2}}, \;E_{C}=-2g\beta(\xi^{2}-1).
\end{equation}
\end{mathletters}
Fluctuations of the atom number in the left well and the phase
difference in this case are given by
\begin{equation}
\Delta n=\frac{1}{\Delta
\phi}=\frac{\sqrt{N}}{\sqrt{2|\xi|}(\xi^{2}-1)^{1/4}}
\end{equation}
where $|\xi|=|g\beta N /(2\gamma)|>1$. In the weak tunneling
regime ($|\xi|\gg 1$), the atom number fluctuation $\delta
n\approx \sqrt{0.5/N}$ is very small in large-N circumstances,
which demonstrates sub-Poissonian fluctuations and the atoms in
any one of the wells can be thought to be approximately in a Fock
state. Another interesting phenomenon is that the atom numbers in
both wells display strong fluctuations when the parameter
$|\xi|\equiv |g\beta N /(2\gamma)|$ approaches one from above. The
atom number fluctuations have the form $\Delta n\approx
\sqrt{N/2}[2(|\xi|-1)]^{-1/4}$ as $|\xi|\rightarrow 1$. This form
demonstrates the typical strong critical fluctuation phenomena in
phase transitions \cite{ker,sac}. The corresponding "order"
parameter and critical index for atom number fluctuation in our
case are $|\xi|$ and $1/4$ respectively \cite{ker,sac}. However,
It should be emphasized that the critical fluctuation phenomenon
corresponds in fact to squeezed states for the phase difference
$\phi$ since the phase difference fluctuation $\Delta \phi\approx
\sqrt{2/N}[2(|\xi|-1)]^{1/4}\rightarrow 0$ as the "order"
parameter $\xi$ approaches unity from above.

The multistability and phase transition as well as the critical
fluctuation phenomenon (i.e., phase squeezed states) in the
macroscopic tunneling in a double-well trap containing a
Bose-Einstein condenstate(BEC) are well within the reach of
nowadays BEC-related technology.  As a matter of fact, the recent
experiment by Kasevich's group\cite{orz} has already reached the
strong tunneling regime and the parameter $\xi=g\beta N
/(2\gamma)$ in that experiment can at least reach as low as 1.5 as
given in the caption of its Fig. 1 D \cite{orz}. We therefore
believe that one should be able to discover the new phenomena
investigated here with the same apparatus as in that experiment.

This work is partially supported by the NSF of
China. Y. Wu acknowledges the hospitality of Key Laboratory of
Quantum Information and Measurements of Chinese Ministry of
Education, Peking University where his part of the work was done.

\end{multicols}

\end{document}